# Electronic excitation of transition metal nitrides by light ions with keV energies

Barbara Bruckner[1,2], Marcus Hans[3], Tomas Nyberg[4], Grzegorz Greczynski[5], Peter Bauer[1,2] and Daniel Primetzhofer[1]

[1] Department of Physics and Astronomy, Uppsala University, 751 20 Uppsala, Sweden

[2] Institute of Experimental Physics – AOP, Johannes-Kepler University Linz, 4040 Linz, Austria

[3] Materials Chemistry, RWTH Aachen University, 52074 Aachen, Germany

[4] Department of Engineering Sciences, Uppsala University, 751 21 Uppsala, Sweden

[5] Thin Film Physics Division, Department of Physics (IFM), Linköping University, 581 83 Linköping, Sweden

E-mail: barbara.bruckner@physics.uu.se

**Abstract**

We investigated the specific electronic energy deposition by protons and He ions with keV energies in different transition metal nitrides of technological interest. Data were obtained from two different time-of-flight ion scattering setups and show excellent agreement. For protons interacting with light nitrides, i.e. TiN, VN and CrN, very similar stopping cross sections per atom were found, which coincide with literature data of $N_2$ gas for primary energies $\leq 25$ keV. In case of the chemically rather similar nitrides with metal constituents from the 5$^{th}$ and 6$^{th}$ period, i.e. ZrN and HfN, the electronic stopping cross sections were measured to exceed what has been observed for molecular $N_2$ gas. For He ions, electronic energy loss in all nitrides was found to be significantly higher compared to the equivalent data of $N_2$ gas. Additionally, deviations from velocity proportionality of the observed specific electronic energy loss are observed. A comparison with predictions from density functional theory for protons and He ions yields a high apparent efficiency of electronic excitations of the target for the latter projectile. These findings are considered to indicate the contributions of additional mechanisms besides electron hole pair excitations, such as electron capture and loss processes of the projectile or promotion of target electrons in atomic collisions.

# 1. Introduction

Ions moving in matter deposit energy due to interaction with either electrons or nuclei of the target, i.e. due to electronic and nuclear stopping. Accurate and detailed knowledge of the energy deposition mechanisms is necessary for fundamental understanding of ion-solid interactions. At the same time it is also decisive for numerous established applications such as sputtering and sputter deposition processes, ion irradiation and implantation, as well as ion beam analysis (IBA) techniques, e.g. for characterization of thin films, surfaces and nanomaterials [1–5]. A measure for the energy loss $dE$ per unit path length $dx$ is the stopping power $S = dE/dx$. To avoid a dependence on the atomic density $n$ of the target, we compare the stopping cross section (SCS) $\varepsilon = 1/n \cdot S$..

For primary energies of at least several hundred keV, energy loss of light ions is mainly due to electronic excitations, i.e. electronic stopping. In this energy regime trajectories are well described within the single scattering model with excitations of inner shells contributing, dependent on the scattering geometry [6]. Towards lower projectile energies, ion-solid interactions are more complex, i.e. screened Coulomb potentials as well as multiple scattering models are required to properly describe energy loss along the trajectory of the projectile. At these low energies both nuclear and electronic stopping contribute to the stopping power, the latter being dominated by excitation of valence electrons as inner shell excitations are limited by the decreasing maximum energy transfer in binary collisions. Consequently, electronic stopping is sensitive to changes in the density of states, as it is the case of compounds. Deviations from Bragg's rule, i.e., additivity of stopping power in a compound according to stoichiometry, are well documented in literature for energies around and below the stopping maximum [7], e.g. in organic compounds [8–10], oxides [11–13], phosphides [14] and silicides [15].

A simple, yet powerful description of target electrons, particularly when considering only delocalized valence and conduction electrons, is a free electron gas (FEG) [16]. In the low energy regime, this approach results in proportionality of the electronic stopping power to the ion velocity, $dE/dx = Q(Z_1, n_e) \cdot v$. Here, $Q$ is equivalent to a friction coefficient, which depends on the atomic number of the ion $Z_1$ and the density of the FEG $n_e$. Usually, the FEG is characterized by the Wigner-Seitz radius, $r_s = \left(3/4 \pi n_e\right)^{1/3}$ (see e.g. [17]), corresponding to a sphere which contains one electron. With the use of nonlinear density functional theory, the ion-electron potentials for a FEG can be calculated, which are used as input parameter to model friction coefficients for different projectiles [18,19]. For slow protons, these DFT calculations are in very good agreement with experiments in manifold target materials by use of effective $r_s$ values [20]. However, in a recent study, clear limitations of the FEG model were reported for proton stopping in rare-earth and early transition metals [21]. For projectiles heavier than protons, even in FEG metals like Al, electronic interactions were found to be more complex [22–24]. Consequently, for He ions deviations from velocity proportionality are reported in literature for many target systems [15,25–27]. Dynamic approaches have in recent years contributed to a deepened understanding of this phenomenon [28–30].

Experimentally, electronic stopping of slow protons in oxides has been studied quite intensively, at energies around the stopping power maximum [31,32], as well as at low proton velocities [13]. Scaling of the SCS with the number of O atoms per building block was observed, with values for $\varepsilon_{\text{O,oxide}}$ that exceed the corresponding values in the gas $\varepsilon_{\text{O}_2,\text{gas}}$ by a factor of ~ 2 [33]. Under the assumption that the electrons are located at the O atoms, the energy loss was attributed to the excitation of the O 2p electrons. For nitrides, below the stopping maximum experimental data are available only for TiN. Here, proton SCS per atom in TiN [27] and $N_2$ gas [34] are found similar as well, resulting in a factor of ~ 2 in the stopping per N atom – in accordance to the observations for oxides and LiF. Again, the number of valence electrons per building block in the nitride and in $N_2$ is similar; TiN is expected to have 7 weakly bound electrons, whereas in $N_2$ gas there are 3 electrons in a N atom. For heavier ions, in the same system a substantially increased energy loss was observed in comparison to predictions for a homogeneous free electron gas [35]. These findings were very recently found to be linked to the contribution of a wide range of electron densities in the system to the observed energy loss [36].

To increase the available understanding of energy deposition by keV ions in nitrides as well as to provide yet unavailable reference data for technologically highly relevant systems we present energy loss measurements for protons and $He^+$ ions in a series of different nitrides, i.e. TiN, VN, CrN, ZrN and HfN. Experiments were performed at energies below the stopping maximum, appropriate for comparison with the above-mentioned models. Furthermore, group IV and V nitrides are employed as refractory compounds and therefore, quantitative predictions are highly relevant for accurate modelling of deposition and materials modification processes such as sputtering and irradiation [37–39]. However, stopping data are scarce for nitrides with virtually no data for transition metal nitrides. Only limited data are available for He in SiN [40], GaN [41] and InN [42] around and above the stopping maximum [34], i.e. energies exceeding the values commonly relevant in ion implantation and sputtering. Of these datasets, in particular InN is of interest, due to significant deviations of the data from SRIM predictions below the stopping maximum.

## 2. Method

Information on the electronic energy loss was deduced from energy spectra of backscattered projectiles. We used two different time-of-flight (ToF) setups to cover a primary energy regime of ~ 1 keV up to ~ 150 keV. In the low energy ion scattering (ToF-LEIS) setup ACOLISSA[43] measurements with protons, deuterium and He ions for primary energies up to 10 keV have been performed at Linz university. This setup features a fixed backscattering angle of $\vartheta = 129°$ with a detector solid angle of $2 \times 10^{-4}$ sr. The medium energy regime was covered with the use of a 350 kV Danfysik ion implanter and a subsequent ToF-medium energy ion scattering (MEIS) beam line [44,45] with a minimum projectile energy of 20 keV and a variable backscattering angle $\vartheta$ in the range from 90° to 160° located at the Tandem Laboratory at the Uppsala University. For the MEIS experiments scattering angles of $155° \pm 2°$ and $160° \pm 2°$ were employed and all measurements – also in the LEIS regime – were performed with a normal angle of incidence. In both setups, protons and molecular ions were used to obtain an overlap in the stopping data between the two laboratories.

We studied two different types of polycrystalline samples: first, nm thin films sputter-deposited on a light substrate (HOPG), and second, thick high-purity bulk nitrides. For the thin films, carbon was chosen as substrate to allow a better disentanglement of the metal peak and the background especially for spectra with lower primary energies.

VN, CrN, ZrN and HfN thin films with intentional thicknesses of 15 and 30 nm were deposited by reactive DC magnetron sputtering in Ar/N$_2$ atmosphere. An elemental V, Cr, Zr and Hf target with 90 mm diameter was assembled in a high vacuum laboratory scale six-way reducing cross and the target-to-substrate distance was 10 cm. The carbon substrates were heated to 400 °C and the base pressure after heating/prior to deposition was always < 2×10$^{-5}$ mbar. N$_2$/(N$_2$+Ar) gas flow ratios of 100 %, 100 %, 20 % and 30 % were employed for the growth of VN, CrN, ZrN and HfN, respectively. A power of 100 W was applied to the cathode at 5×10$^{-3}$ mbar deposition pressure and the substrate bias potential was floating. The deposition time was adjusted accordingly based on reference synthesis experiments for 60 minutes combined with cross-section scanning electron microscopy thickness measurements. The venting temperature was < 40 °C for all synthesis experiments in order to minimize oxygen incorporation during exposure to atmosphere [46].

The highly crystalline TiN film was sputtered in a cylindrical UHV vacuum chamber (Kurt J. Lesker CMS-18) equipped with a load-lock and evacuated by a cryopump (CTI CryoTorr 8). The base pressure was below 10$^{-7}$ mbar. The substrate table was rotated at 20 rpm and kept at floating potential. The pulsed DC power to the target was supplied by an Advanced Energy Pinnacle Plus power supply. Pure nitrogen (30 sccm) was introduced in the chamber and the processing pressure was 1.2×10$^{-3}$ mbar. The substrate temperature was kept at 750 °C and the 4 inch target was fed with 800 W of pulsed DC (resulting in a target current of 1.35 A) with a frequency of 50 kHz and a pulse off-time of 0.5 µs.

The thicker, polycrystalline ZrN, and HfN film layers were grown in an industrial deposition system [47] employing rectangular 8.8×50 cm$^2$ elemental targets by reactive DC magnetron sputtering (DCMS) in Ar/N$_2$ gas mixtures at the total pressure of 4×10$^{-3}$ mbar. The N$_2$-to-Ar flow ratio was optimized for each target material to obtain stoichiometric single-phase layers. Films were grown on Si(001) substrates sequentially cleaned in acetone and isopropanol and mounted 18 cm away from the target surface. The average target power was 2 kW, while the substrate bias was set at -60 V DC. Two resistive heaters operating during film growth at 8.8 kW each, ensure the substrate temperature of (470 ± 12) °C. All films are exposed to the laboratory atmosphere at very similar venting temperatures $T_v$ = (230 ± 20) °C in order to avoid significant influence of $T_v$ on the thickness of formed surface oxide layer [46]. More details related to sample deposition as well as thorough characterization are reported in Ref. [48].

All samples were characterized with the use of Rutherford Backscattering Spectrometry and ToF-Elastic Recoil Detection Analysis (ERDA) at the Tandem Laboratory to evaluate their stoichiometry, purity and areal density. An extensive discussion of the setups and typically employed procedures for data evaluation can be found elsewhere [49]. The thickness measured as areal density was converted into Å via the bulk densities for the nitrides [50]. The resulting thicknesses are in good agreement with physical thickness measurements by scanning transmission electron microscopy, which confirms the density of the films being close to bulk materials. In Tab. 1 we show an overview of composition as well as thickness of the investigated samples as obtained from RBS and ERDA. All of the thin films exhibit low Z impurities mainly O, but also small amounts of C or Ar from the sputtering process with concentrations for all impurities < 13 %. In Hf and Zr, the respective other metal was found as impurity in all samples (due to their chemical similarity, Hf and Zr are difficult to separate).

In RBS normalization of the charge solid angle product with respect to simulations (SIMNRA [51]) can be performed by either using the intensity of the yield in the plateau originating from the carbon substrate or by a relative measurement of the sample to e.g. Au, Ag or Cu bulk films. In both cases, the electronic stopping in these materials has to be well-known [52]. However, for C the scatter in published electronic stopping data at ~ 400 keV is larger than for Au at ~ 2 MeV. The different relevant energies result from the large difference in the corresponding kinematic factors for C and Au. For measurements relative to Au or Cu a systematic uncertainty of < 2 % from the electronic SCS has to be considered [53]. For the determination of the nitrogen content a statistical uncertainty of ~ 4 %, due to the background in the RBS spectra originating from impurities in the C-substrate and pile-up in the detector has to be considered.

|  | thickness [Å] | N/TM | impurities [%] | C [%] | O [%] | Ar [%] | Zr / Hf [%] |
|---|---|---|---|---|---|---|---|
| HfN | 132 | 1.27 | 7 | - | 2.5 | 2 | 2.5 |
|  | 262 | 1.38 | 6 | - | 2 | 1.5 | 2.5 |
|  | thick [48] | 1.265 |  |  |  |  |  |
| ZrN | 133 | 1.20 | 10 | - | 10 | 0.5[1] | 0.2[1] |
|  | 277 | 1.20 | 5 | - | 5 | 0.5[1] | 0.2[1] |
|  | thick [48] | 1.00 |  |  |  |  |  |
| CrN | 149 | 1.02 | 8 | 1.8 | 6.2 |  |  |
|  | 300 | 1.08 | 7.1 | 1.8 | 5.3 |  |  |
| VN | 304 | 1.15 | 12.7 | 4.6 | 8.1 |  |  |
| TiN | 85 | 1.09 | 8 | - | 8 |  |  |

Table 1: Sample thicknesses and composition of the investigated transition metal nitrides. All thin films were grown on HOPG. Typically, for nitrides we give the ratio of nitrogen to metal constituents (N/TM).

At the employed low- and medium- primary energies, a comparison of the spectra of backscattered projectiles to Monte-Carlo simulations is necessary to disentangle electronic stopping from effects of multiple scattering and nuclear stopping. For this purpose, we used the TRIM for backscattering code (TRBS), which allows adjustments in the interatomic potential, affecting contributions from multiple scattering, and for electronic stopping along the trajectories. In order to evaluate electronic stopping from the measured energy spectra, $\varepsilon$ was varied in the simulation to obtain optimum agreement between the relevant features in experiment and simulation. Note, that this procedure yields a mean SCS that is averaged over all (possibly impact parameter dependent) electronic energy loss processes in the target.

In the TRBS simulations, we employ the Ziegler-Biersack-Littmark (ZBL) potential [54], which in general shows good agreement between the experimental spectra and the TRBS simulation. However, for HfN deviations in the multiple scattering background are observed for $E_0 \leq 100$ keV. At lower primary energies screening length corrections are needed to reproduce the background in the experiment. A recent study investigates the influence of uncertainties in the scattering potential on the evaluation of electronic energy loss. As a result, the evaluation of the peak width for energies $E_0 \leq 100$ keV may require differences in electronic stopping, necessary to obtain a good fit, up to ~ 3 % depending on the employed potential – ZBL or Thomas-Fermi-Moliere – and a possible screening correction [55].

---

[1] not considered for the evaluation of $\varepsilon$

For thin films, electronic stopping can be deduced from the width of the metal peak, if film composition and thickness are known [56]. Figure 1 shows typical time-of-flight spectra converted to energy obtained with the (a) ToF-LEIS and (b) ToF-MEIS setup. Panel (a) depicts the spectrum for 8 keV protons scattered from a TiN film with a nominal thickness of 85 Å on a carbon substrate; in panel (b) the spectrum for 60 keV He$^+$ ions backscattered from a 149 Å thin CrN film on carbon is plotted for a scattering angle of $\vartheta = 155° \pm 2°$. The resolution of the trailing edge in the experimental spectrum depends on the energy resolution of the primary ion beam, multiple and dual scattering, energy loss straggling and the energy resolution of the detection system. Variations in the thickness of the sample are not expected to contribute significantly to the trailing edge, as at higher energies good agreement between experiment and the simulation is found. Figure 1 also includes TRBS simulations with a best fit to the experiment (red solid line) as well as modifications of $\varepsilon$ by $\pm$ 10 % and $\pm$ 5 % (red dashed lines). In the TRBS code multiple scattering as well as increases in path length are sufficiently implemented, however electronic energy loss straggling is not, but is by definition not expected to alter the mean energy loss. Therefore, the width in the trailing edge of the simulation corresponds exclusively to influences from multiple scattering as well as the energy resolution from the experiment. For energies as low as 8 keV H$^+$ the signals originating from the thin film and the substrate can still be disentangled as depicted in Fig. 1(a) and therefore, the substrate signal is not expected to influence the evaluation of the energy loss. At lower energies, where the signals originating from the metal constituent and from the backing cannot be clearly separated, information on the energy loss in the film can be obtained from the height of the signal as described below.

Note that, in any approach, the stopping deduced from the simulation corresponds to all electronic losses experienced in the film including impurities. We therefore, performed a correction according to Bragg's rule [57] to obtain the electronic SCS in the nitride, as described in Ref. [58]. This correction in the SCS amounts to at most 6 %, with largest impact on the low energy data.

For the thick samples, electronic stopping can be evaluated from a comparison of the intensity of the spectrum at an energy slightly below the high-energy onset of the metal constituent to the corresponding spectrum height of a reference material at energies close to but below the leading edge of the spectrum [52,56]. In this way, the influence of possible surface impurities is excluded. Note, here the electronic stopping in the reference material $\varepsilon_{ref}$ must be known in order to obtain SCS data for the sample of interest $\varepsilon_x$. Consequently, uncertainties in $\varepsilon_{ref}$ yield a systematic error in the evaluated $\varepsilon_x$. In general, for these measurements a reference with similar atomic number as the sample of interest is chosen in order to minimize the influence of uncertainties in the scattering potential on data evaluation. Therefore, Au [59] and Ag [60] have been used as reference for HfN and ZrN, respectively.

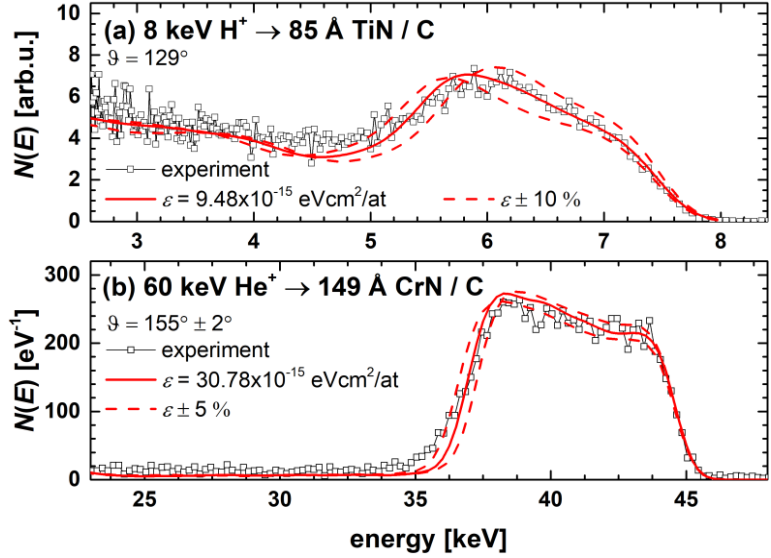

Figure 1: ToF converted energy spectra for (a) 8 keV H$^+$ scattered from TiN and (b) 60 keV He$^+$ ions scattered from CrN. The experimental data (black squares) are compared to TRBS simulations with different electronic stopping values: a best fit and simulations with variations in the electronic stopping of (a) ± 10 % and (b) ± 5 % are depicted as red solid and dashed lines, respectively.

## 3. Results and discussion

The electronic stopping data of protons in five different nitrides are plotted in Fig. 2; the upper panel (a) depicts data for three nitrides with metal constituents from period 4 and an increasing number of valence electrons: TiN (black squares), VN (dark blue pentagons) and CrN (magenta hexagons). Data from literature for TiN [27] are plotted as grey triangles; no data is available in this energy regime for the other systems. Excellent agreement between the present measurements and data from literature can be observed. The SCS data for VN and CrN coincide in the investigated energy regime within experimental uncertainties, and are slightly lower than data for TiN.

In panel (b) of Fig. 2 we show energy loss data for three nitrides with the same number of valence electrons, i.e. based on group IV transition metals: TiN (black squares and grey triangles [27]), ZrN (red points) and HfN (green diamonds). Where recorded (HfN, ZrN), data obtained from both evaluation methods and both setups agree within their experimental uncertainties. For all data in Fig. 2 the systematic uncertainties originating from sample characterization as well as the employed scattering potential are indicated for the highest and lowest investigated energy. The statistical uncertainty can be obtained from the scatter of the energy loss data itself. Due to the lack of data for most of the nitrides we compare our results at first hand to predictions by Bragg's rule [57] using SRIM-13 [61] data for the constituents. The obtained values coincide with direct predictions by SRIM-13 for the corresponding nitrides and are shown as dashed lines in Fig. 2 with the same color code as the experimental data. The predictions do not quantitatively describe stopping in this energy regime, with deviations up to ~ 22 % observed. While SRIM generally overestimates the energy deposited, predictions reproduce, however, the correct order of the magnitude of the experimental data: ZrN exhibits highest and CrN and VN lowest stopping cross sections. Based on the principles of SRIM these observations can be understood since for some metals, as e.g. Hf and Zr, either no or only scarce experimental data are available, especially below the stopping maximum [34]. In these cases, SRIM stopping is based on interpolations from other metals. For samples of different stoichiometry for the same compound, as e.g. HfN, only one prediction by Bragg's rule is depicted in Fig. 2, since the small difference in the SCS resulting from different stoichiometries (see Tab. 1) could not be resolved within experimental uncertainties.

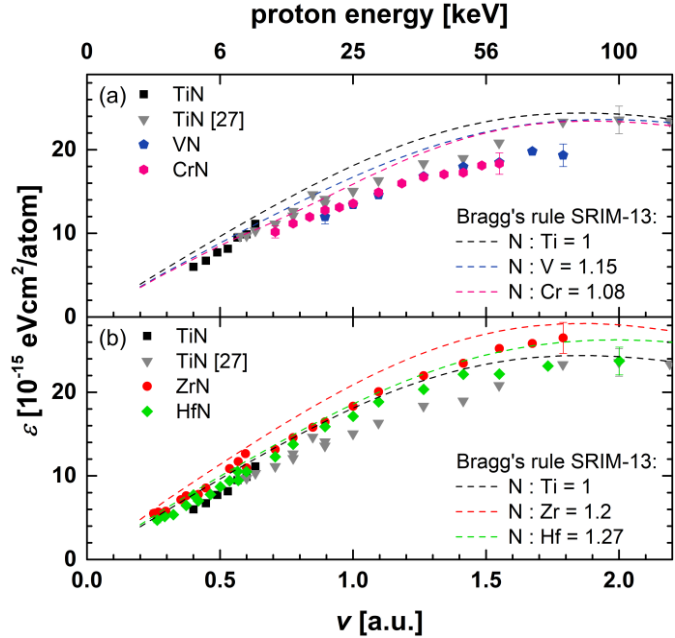

Figure 1: Electronic stopping cross sections for protons in (a) TiN (black squares), VN (dark blue pentagons) and CrN (magenta hexagons) as well as (b) TiN (black squares), ZrN (red points) and HfN (green diamonds). Additionally, data from literature are plotted for TiN [27] in both panels. The different dashed lines correspond to predictions by Bragg's rule based on data by SRIM-13 in the color code of the corresponding materials.

To simplify comparison with predictions by theory, in Fig. 3, we plot data normalized to the ion velocity, again with the same color code as in Fig. 2. Additionally, also data for $N_2$ gas from literature [34] are depicted as blue asterisks. For protons and $v$ below ~ 1 a.u., $\varepsilon$ exhibits velocity proportionality as predicted for a FEG [16], with different slopes for different nitrides. For velocities up to ~ 1.0 a.u. (indicated as dashed line in Fig. 3), the $\varepsilon/(\text{atom} \times \text{a.u.})$ values for VN and CrN coincide with $\varepsilon/(\text{atom} \times \text{a.u.})$ in $N_2$ gas within experimental uncertainties and data for TiN only slightly exceed the values for $N_2$. Thus, in the investigated nitrides the SCS per building block is found equal to the value per $N_2$ molecule in the gas. In other words, the nitride SCS per N atom is twice as high as it is in the gas phase. This observation is in striking analogy to the oxides, for which the SCS/O atom is found twice as high as for $O_2$ gas [13].

In the following section, we will discuss the observed effects from two distinct perspectives, i.e. an atomistic one, as well as employing a free electron gas model. This approach is made, as nitrides are found to feature a wide range of properties, similarly as for oxides. While they formally have an oxidation state of -3, justifying the atomistic picture, they also feature good electrical conductivity, which justifies an electron gas approach. In an atomistic picture, the present data for protons could be naïvely understood by complete electron transfer. Assuming this transfer would result in an electron configuration of N $2p^6$, which would be twice as efficient in electronic stopping as N $2p^3$, ignoring changes in binding energy and their impact on excitation. Such an approach should be appropriate at least for strongly ionic compounds like Alkali halides, where valence electrons are clearly located at the anions [62]. Indeed, for LiF [63,64], $\varepsilon$/atom for protons was observed to be similar to experimental data for Ne [65] and to expectations for F [66]. Equivalently, comparison of $\varepsilon_{LiF}$/F atom LiF to $\varepsilon_{Ne}$ and the prediction for $\varepsilon_{F,gas}$ yields $\varepsilon_{LiF}$/F atom $\approx 2\varepsilon_{Ne} \approx 2\varepsilon_{F,gas}$ for 10 keV protons. There are, however, major differences in the electronic structures of fluorides, oxides and nitrides: While for alkali halides and to a lower extend in the oxides the density of valence electrons is dominantly centered at the anions, in nitrides the valence electron densities are comparable at the metal sites and at N [62]. For the oxides studied in [13], no significant evidence for an influence of the different band gaps (0 to 10 eV) was observed. Thus, we expect the observed simple scaling for nitrides to be the result of a complex interplay of modifications of the electronic structure when comparing the more ionic, anion-centered electronic structure to a more covalent bond as in the nitrides together with drastically changing ionization energies.

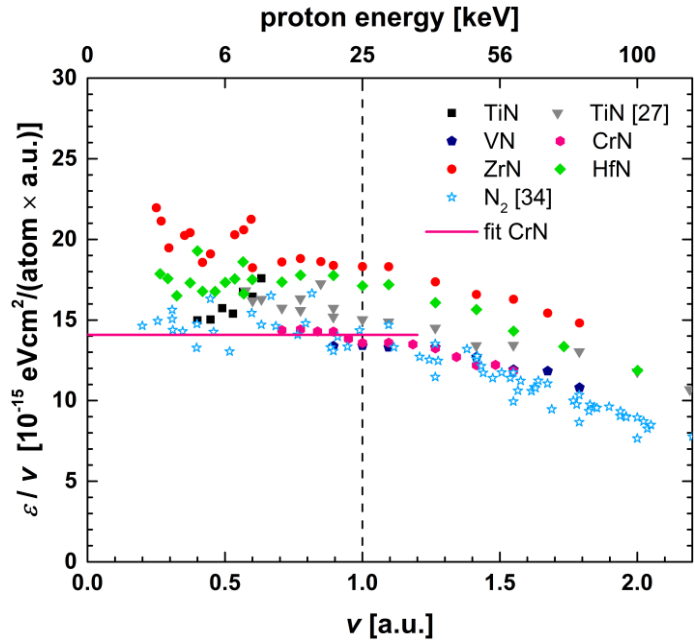

Figure 3: Electronic stopping cross sections per ion velocity for protons in the different transition metal nitrides: TiN (black squares), VN (dark blue pentagons), CrN (magenta hexagons), ZrN (red points) and HfN (green diamonds). Additionally, data from literature are plotted for TiN [27] and $N_2$ (gas) [34]. A fit to data for CrN is depicted as magenta solid line resulting in an effective density of the FEG of $r_{s,eff}$ = 1.62 a.u. according to DFT calculations [67]. For velocities below ~ 1 a.u. (dashed black line) electronic stopping is expected to be velocity proportional.

A completely different approach is to compare the present data to predictions for a FEG as obtained from DFT. In such an approach, a fit to the CrN data (magenta line in Fig. 3) yields a friction coefficient of $Q_H = 0.294$ a.u. resulting in an effective electron gas density of $r_{s,eff} = 1.62$ a.u. according to DFT calculations [67]. This density would correspond to ~ 7 electrons contributing to electronic stopping for CrN as well as VN and TiN, due to their similar SCS, in reasonably good agreement with expectations from plasmon frequencies [68–70]. However, we are well aware that the unperturbed density of states for the investigated compounds is fundamentally different in structure from a FEG, with occupied and unoccupied electronic states featuring high densities in narrow energy bands. Also, the comparable velocities of projectiles and electrons involved in the excitation process may require a more complex dynamic description. For example, in a recent work, the free electron gas model has been shown to break down for early transition metals from the 6$^{th}$ period [21]. At the same time, the magnitude of the electronic energy loss of slow ions in ionic crystals such as LiF could be successfully reproduced in this framework [71]. A recent work by Matias et al. [36] showed, how similar results for protons can be obtained by strongly spatially dependent electron densities, i.e. in a non-homogeneous electron system.

In the present investigation, for the systems with more complex electronic structure, i.e. ZrN and HfN, the SCS is significantly higher than for TiN, VN and CrN, with highest SCS for ZrN. Based on the conclusions from above, the real nature of excitation is thus considered to depend on details of electronic structure, in particular its local character [36], with higher local electron densities at the cations for Zr and Hf.

In Fig. 4 SCS data for He ions are presented similarly as for protons, i.e. (a) TiN (grey triangles) [27], VN (dark blue pentagons), CrN (magenta hexagons) and (b) TiN (grey triangles) [27], ZrN (red circles) as well as HfN (green diamonds). Analogue to protons the systematic uncertainties for each sample are plotted only for the lowest and highest energy. Again, data for TiN, VN and CrN coincide within the experimental uncertainties, whereas the specific energy deposition in ZrN and HfN is higher. We also show SRIM predictions according to Bragg's rule, due to the lack of existing experimental data for the other compounds. Similarly as for protons, the order of the predicted values is the same as we observe with our experimental data, however, with significant discrepancies in the absolute values. Specifically, for the present projectile SRIM is found to underestimate the actual energy deposition by up to 25%.

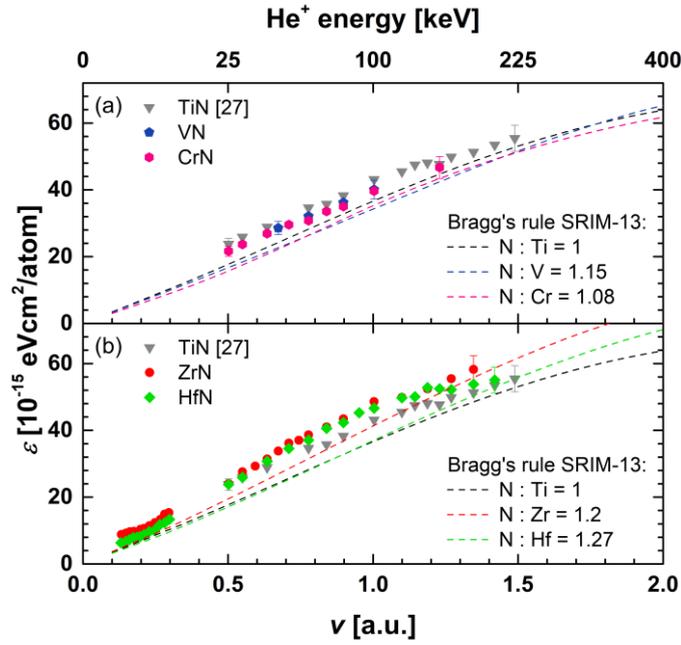

Figure 4: Electronic stopping cross section helium ions in (a) TiN (grey triangles) [27], VN (dark blue pentagons) and CrN (magenta hexagons) as well as (b) TiN (grey triangles) [27], ZrN (red points) and HfN (green diamonds). The different dashed lines correspond to electronic stopping predicted by SRIM according to Bragg's rule.

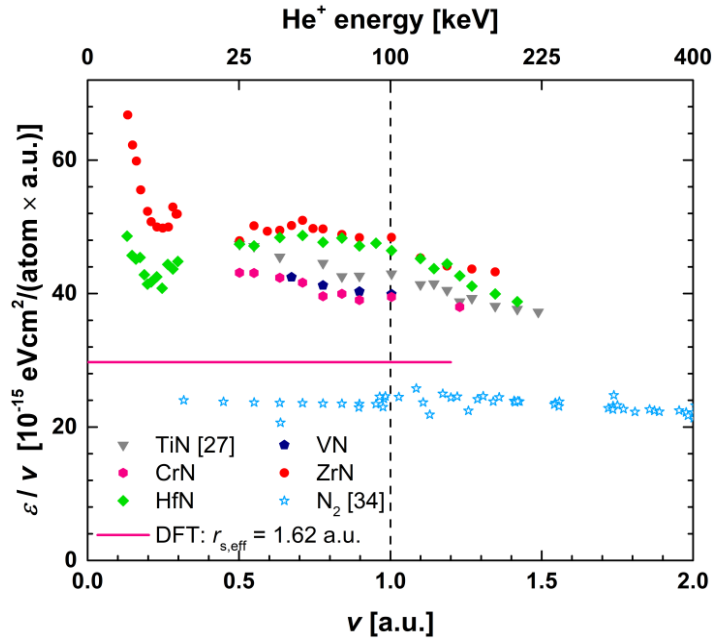

Figure 5: Stopping cross sections per He ion velocity for the different nitrides: TiN (grey triangles) [27], VN (dark blue pentagons), CrN (magenta hexagons), ZrN (red points) and HfN (green diamonds). Additionally, data for $N_2$ gas are depicted from literature [34]. The magenta line would correspond to the electronic energy loss as predicted from DFT assuming the same density for the FEG as obtained for protons.

Figure 5 depicts again the electronic stopping cross section for He normalized by the velocity for all investigated nitrides: VN (dark blue pentagons), CrN (magenta hexagons), ZrN (red points) and HfN (green diamonds). Additionally, data from literature for TiN (grey triangles) [27] as well as $N_2$ gas [34] are plotted. In contrast to protons, in all nitrides, the $\varepsilon/(\text{atom} \times \text{a.u.})$ for He ions is significantly higher than in $N_2$, by up to a factor of two for low energies, in clear contrast to expectations from a naïve atomistic model. A similar discrepancy between data for protons and He has been observed for other target materials with vastly different electronic structures [23,72,73]. On the same line, also DFT-based predictions for the same electron gas density as employed to model our data for protons (magenta line in Fig. 5) underestimate the energy loss of He ions. Note, in this context, that the above-referred recent work by Matias et al. correctly reproduces the ratio of the electronic stopping powers for protons and He in TiN based on correct inhomogeneous electron densities, and thus should be employed to model also the observations made for the present systems.

In a more detailed analysis of the velocity dependence, data for TiN, VN and CrN show an apparent threshold when extrapolating medium energy data for He stopping towards zero velocity. In ZrN and HfN, $\varepsilon_{He}$ is proportional to the ion velocity for 0.2 a.u. $< v <$ 1 a.u. However, at velocities below ~ 0.2 a.u. it deviates from this proportionality and extrapolation towards $v \rightarrow 0$ leads to a positive offset. While such an offset value is per se meaningless; it is, however, well in agreement with the presence of an additional energy loss process, with a different and weak velocity dependence.

## 4. Summary and conclusions

We presented data for the specific electronic energy loss of H and He ions in different transition metal nitrides at energies from a few to 200 keV. Samples were synthesized following different routines, and with different physical structure i.e. significantly different thickness. Together with extensive characterization this procedure was conducted to minimize a possible influence of sample impurities and microstructure. Stopping cross sections were obtained from evaluating the width of resulting energy spectra. For thicker films, stopping cross sections were deduced from fitting the spectral intensity. In this case electronic stopping was evaluated relative to a reference material of accurately known stopping power, i.e. Ag and Au. Where both methods have been applied, good agreement of the deduced datasets was observed.

From an applied point of view, for protons, electronic energy deposition was found to be systematically lower than predictions by SRIM. On the contrary, for He, data was found high with respect to SRIM, with maximum discrepancies up to 25%. The relative magnitude for the different compounds, however, was found in accordance to predictions. From a fundamental perspective, for protons electronic energy loss below ~ 1 a.u. (25 keV) in the light transition metal nitrides, i.e. TiN, VN and CrN, was found to be a factor of ~ 2 more effective than in molecular $N_2$ gas, similar to observations for oxides [13]. However, for ZrN and HfN electronic energy loss is significantly higher. For He ions, $\varepsilon$ in the nitrides is more effective than expected exclusively from electron-hole pair excitations. The observed deviations from velocity proportionality of the electronic stopping for He ions indicate additional local contributions to the energy loss rooted in non-homogeneous electron densities and additionally possible enabling excitation channels such as electron capture and loss between projectile and target, which alter effective charge states. These phenomena are expected to severely affect the local character of energy deposition, the secondary electron spectrum and thus radiation effects such as defect creation.

The obtained data are expected to act as benchmark systems for advanced possibly time-dependent modelling of electron dynamics for inhomogeneous systems, which can adequately describe such phenomena to confirm or reject this hypothesis. At the same time, data are expected to contribute to more reliable semi-empirical modelling of the energy deposition and thus accurate modelling of processes such as sputtering or ion implantation.

**Acknowledgment:**

Financial support of this work by the Swedish Foundation for Strategic Research SSF in the form of an infrastructure fellowship (RIF14-0053) for the accelerator facility and the Austrian Science Fund FWF (FWF-Project No. P22587-N20) is gratefully acknowledged. BB is also grateful to the Wilhelm-Macke Foundation at the JKU for supporting her stay at Uppsala University. PB expresses his gratitude for the kind hospitality at UU. GG acknowledges the financial support from the Knut and Alice Wallenberg Foundation Scholar Grant KAW2016.0358, the Swedish Research Council VR Grant 2018-03957, the VINNOVA Grant 2018-04290, and Carl Tryggers Stiftelse Contract CTS 17:166.

**Conflicts of interest**

There are no conflicts to declare.